\documentclass[useAMS,usenatbib]{mn2e}

\usepackage{graphics}
\usepackage{epsfig}

\title[Contact binaries at high galactic latitudes]
{Luminosity function of contact binaries at high 
galactic latitudes towards the LMC and the SMC}

\author[Pascal D. Nef and Slavek M. Rucinski]
{Pascal D. Nef$^1$, Slavek M. Rucinski$^2$ \\
$^1$Swiss Federal Institute of Technology (ETH-Zurich), 8093 Zurich, Switzerland; 
pnef@phys.ethz.ch\\
$^2$Department of Astronomy and Astrophysics, University of Toronto, \\
50 St.~George St., Toronto, Ontario M5S~3H4, Canada; rucinski@astro.utoronto.ca}

\date{Accepted --.      Received -- ;      in original form --}

\pubyear{2007}

\begin{document}
\label{firstpage}
\maketitle

\begin{abstract}
Using the OGLE catalogue of eclipsing binaries, 15 contact binaries 
were identified towards the SMC and the LMC at vertical distances 
from the Galactic plane between 300 pc and 10 kpc. 
Based on the luminosity function calculated for these contact 
binaries, we estimated a frequency of occurrence relative to 
Main Sequence stars in the thick disk at roughly $\frac{1}{600}$. 
This estimate suffers from the small number statistics, but is consistent 
with the value previously found for the solar neighbourhood.
\end{abstract}


\section{Introduction}

During the past few decades, several attempts have been made 
to determine the local spatial density of contact binaries, 
with very diversified results. \citet{ru02} (where full
references are given and differences discussed) estimated their local 
spatial density at $(1.02\pm0.24) \times 10^{-5} \ \mathrm{pc^{-3}}$.
Later, on the basis of the All Sky Automated Survey (ASAS), 
a relative frequency of occurrence (RFO) of one contact binary 
among about 500 solar type stars was derived \citep{ru06},
which is in full agreement with the mentioned spatial density. 
The RFO for contact binaries at high galactic latitudes, however, 
has been entirely unknown. 

Here we present an estimate for 
the spatial occurrence of contact binary systems relative 
to Main Sequence (MS) stars in the thick disk of our Galaxy, far
from the Galactic plane. In particular, the luminosity function 
for contact binaries in two conical volumes towards the Small Magellanic 
Cloud (SMC) and the Large Magellanic Cloud (LMC) covering parts of the 
thick disk and the halo is determined. The results are then 
compared with the luminosity function for MS stars in the 
same region of the sky. The term ``contact binaries'' is here 
used as a synonym for W~UMa-type eclipsing binaries with 
orbital periods in a range of 0.22 -- 1 days. In this paper,
for reasons to be explained below, we limit ourselves to a subset
with periods $<0.45$ days.

\section[Identification of contact binaries]%
	{Identification of short period contact binaries in 
       the OGLE-catalogue of eclipsing binaries}

The Optical Gravitational Lensing Experiment (OGLE) was 
intended to detect dark matter in the Milky Way Galaxy 
using the microlensing technique, with the Magellanic Clouds 
and the Galactic Bulge being the main targets of the 
survey \citep{usk97}. As a by-product, OGLE provides high quality, 
long-term photometry that can be used 
to analyse eclipsing binary stars. Two online 
catalogues \citep{wy03,wy04} were used for this study. 
They contain data in the standard 
photometric $BVI$ system for 2850 and 1351 eclipsing binaries, 
and cover an area of 4.6 and 2.4 square degrees of the central 
parts of the LMC and the SMC respectively. The OGLE-II survey 
has a faint limit of roughly $I=20.5 \ \mathrm{mag}$ with a 
corresponding error of $0.3 \ \mathrm{mag}$; the error 
becomes smaller for decreasing magnitude, reaching a bright 
limit of the survey at about 13 mag \citep{wy04}.

\begin{figure*}
\begin{center}
\includegraphics[width=176mm]{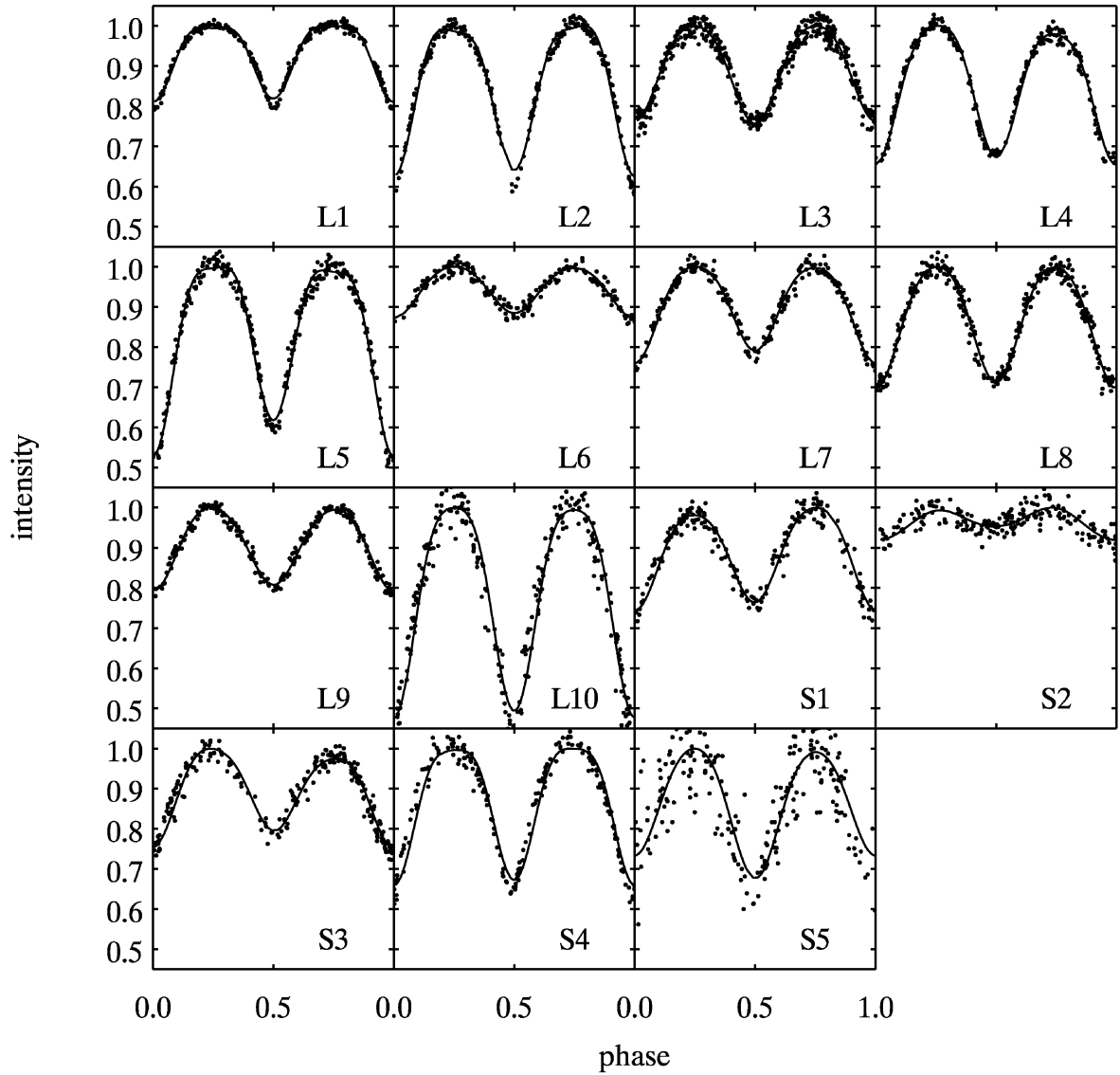} 
\caption{\label{fig_multi_plot} The light curves of the 15 close contact binaries. 
The fitting was done according to the approach of \citep{ru93} 
(i.e. using only the first 5 cosine and the first sine Fourier-coefficients).}
\end{center}
\end{figure*}

The differentiation between contact binaries and other 
eclipsing binary types in the two OGLE catalogues was 
carried out by applying a contact binary criterion based on 
Fourier analysis of the light curves, which was introduced 
by \citet{ru97}\footnote{More detailed explanations for this 
and other techniques used in this paper can be found in \citep{ru97} 
and \citep{ru06}.}. By means of this criterion and using 
a visual inspection of the remaining light curves, 
we identified 10 and 5 contact binaries with orbital 
periods $P<0.45 \ \mathrm{d}$ towards the LMC and the SMC,  
respectively. The light curves are plotted in Figure \ref{fig_multi_plot}.
Furthermore, we used the online database 
from the MACHO (Massive Compact Halo Object) project \citep{al97} 
to confirm the classification of the contact binaries 
in the OGLE catalogue. A direct comparison with the OGLE catalogue 
was only possible for the contact binaries towards the LMC, 
because there is no MACHO photometry available for most 
survey fields covering the SMC. The MACHO photometry of the 10 contact binaries
towards the LMC is in good agreement with the results obtained from 
the OGLE survey\footnote{As an exception, the OGLE photometry 
in $V$-band for the contact binary OGLE050905.22-693315.1 
was found to have a gross error of roughly 3 magnitudes.}.

The contact binaries in our sample have orbital periods in a range of
0.22 -- 0.45 days. The short period limit 
is a natural cut-off for contact binaries \citep{ru07},
whereas the upper period limit was intentional: On one hand, we wanted 
to be sure of the contact binary classification and avoid 
semi-detached binaries, while on the other hand, 
we wanted to use the simple $M_V \equiv M_V (\log P)$ 
calibration and avoid problems with uncertain or missing 
colour indices. Because the final RFO estimate is done using 
absolute magnitude bins of the luminosity function, 
the period limits signify an intentional restriction to the low brightness
end of the contact binary sequence. We will explain the details below.

\section{Absolute magnitudes \& distances}
\begin{figure}
\begin{center}
\includegraphics[width=84mm]{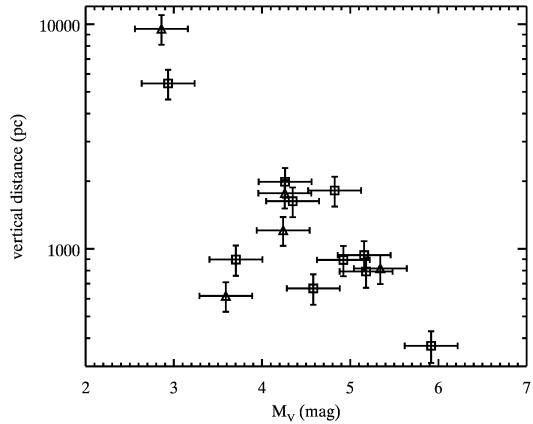}
\caption{\label{fig-cb_dist} This plot shows the vertical distance 
from the galactic mid-plane for the 10 and 5 contact binaries 
towards the LMC (squares) and the SMC (triangles) respectively. 
The Sun is assumed to be located $30 \ \mathrm{pc}$ above 
the galactic mid-plane.}
\end{center}
\end{figure}

Contact binaries show a correlation between the 
colour (i.e. $(B-V)_0$) and the orbital period $P$, 
which was first observed by \citet{eg67}.
As was shown by \citet{RD1997}, this 
correlation can be used to determine an absolute 
magnitude calibration $M_V = M_V(\log P, (B-V)_0)$; an 
even simpler version, $M_V = M_V(\log P)$, is applicable
\citep{ru06} when a restriction to short periods is
added. In this research, the 
absolute magnitude calibration 
$M_V=-1.5(\pm 0.8)-12.0(\pm 2.0) \log P$ was used, 
which is applicable for $P<0.562 \ \mathrm{d}$ \citep{ru06}. 
While applying this calibration to the contact binary sample, 
we arbitrarily assumed an uncertainty in absolute 
magnitude of $0.3 \ \mathrm{mag}$; the formal 
uncertainties in the calibration were larger and
probably overestimated. 
Using this absolute magnitude calibration 
and the relation between the distance and the 
apparent and absolute magnitudes, 
$d={\rm dex}{(\frac{V-M_V+5-A_V}{5})}$, 
the distance from the sun 
and the galactic height for the 15 contact binaries 
were estimated. The extinction towards the LMC and SMC 
was obtained from \citep{sc98}; we assumed $A_V=3.1 E_{B-V}$. 
Figure \ref{fig-cb_dist} illustrates the vertical distance 
from the galactic mid-plane to the 15 contact binaries as a 
function of the absolute magnitude, whereas 
Table~\ref{CB-data-table} lists the numerical results.

% plot M_V, dist for LMC & SMC

% table listing name, V, M_V, P, d, d_z
\begin{table*}
\begin{minipage}{164mm}
\caption{The short period contact binaries towards the LMC and the SMC. 
The table lists the name of the star (OGLE convention),
the Magellanic Cloud field, the orbital period, 
the apparent magnitude in $V$-band, the absolute
magnitude $M_V$, the distance along the line of sight and 
the vertical distance above the Galactic mid-plane. }
\label{CB-data-table}	
\begin{tabular}{@{}ccccccc}
name & field & period $(d)$ & $V$ $(mag)$ & $M_V$ $(mag)$ & distance $(kpc)$ & vertical distance $(kpc)$ \\
\hline 
OGLE003835.24-735413.2 (S1) & SMC & 0.26909 & 15.89 & 5.3 & $1.22 \pm 0.17$ & $0.82 \pm 0.12$  \\
OGLE005431.85-723510.9 (S2) & SMC & 0.33125 & 16.44 & 4.3 & $2.59 \pm 0.36$ & $1.77 \pm 0.26$ \\
OGLE005846.47-724315.3 (S3) & SMC & 0.33244 & 15.61 & 4.2 & $1.78 \pm 0.25$ & $1.21 \pm 0.18$ \\
OGLE004619.65-725056.2 (S4) & SMC & 0.37663 & 13.55 & 3.6 & $0.93 \pm 0.13$ & $0.62 \pm 0.09$ \\
OGLE004849.85-725554.8 (S5) & SMC & 0.43328 & 18.67 & 2.9 & $13.78 \pm2.01$ & $9.54 \pm 1.44$ \\
OGLE054003.85-703837.3 (L1) & LMC & 0.24091 & 15.48 & 5.9 & $0.73 \pm 0.10$ & $0.37 \pm 0.06$ \\
OGLE050542.01-691725.9 (L2) & LMC & 0.27755 & 16.31 & 5.2 & $1.51 \pm 0.21$ & $0.79 \pm 0.12$ \\
OGLE051932.28-694633.4 (L3) & LMC & 0.27871 & 16.64 & 5.2 & $1.78 \pm 0.25$ & $0.94 \pm 0.14$ \\
OGLE053540.37-695413.9 (L4) & LMC & 0.29159 & 16.30 & 4.9 & $1.69 \pm 0.23$ & $0.89 \pm 0.14$ \\
OGLE053251.73-700256.2 (L5) & LMC & 0.29718 & 17.71 & 4.8 & $3.39 \pm 0.47$ & $1.82 \pm 0.28$ \\
OGLE053916.98-700903.3 (L6) & LMC & 0.31132 & 15.35 & 4.6 & $1.28 \pm 0.18$ & $0.67 \pm 0.10$ \\
OGLE050905.22-693315.1 (L7) & LMC & 0.32563 & 17.00 & 4.3 & $3.05 \pm 0.42$ & $ 1.63 \pm 0.25 $ \\
OGLE053247.54-694403.1 (L8) & LMC & 0.33104 & 17.34 & 4.3 & $3.71 \pm 0.51$ & $1.99 \pm 0.30$ \\
OGLE053539.86-694759.5 (L9) & LMC & 0.36844 & 15.09 & 3.7 & $1.70 \pm 0.24$ & $0.90 \pm 0.14$ \\
OGLE054701.27-705623.3 (L10) & LMC & 0.42700 & 18.18 & 2.9 & $10.01\pm 1.41$ & $5.44 \pm 0.82$  \\
\end{tabular}
\end{minipage}
\end{table*}

\section[Luminosity Function]{The amplitude distribution and the
luminosity function for the 15 contact binaries}
Small-amplitude systems often remain undetected in variable 
star searches when the photometric error is too large for detection 
of a small photometric variability. The number of 
missed small amplitude systems was estimated by comparing 
our sample with a theoretical model of the amplitude 
distribution for contact binaries \citep{ru01}. This model predicts 
the amplitude distribution as a function of the degree
of contact $f$ and the mass ratio of the two companions $q=\frac{M_2}{M_1}$
for a sample of contact binaries. 
In this research, we assumed $f=0.25$, what appears to be the favoured 
degree of contact, and a flat distribution of $q$.
We applied this model in the same way as in \citep{ru06}, where more 
information can be found. Following this approach,
we multiplied the number of contact binaries by factors of
1.6 and 1.5 for the LMC and the SMC respectively; both 
factors have an uncertainty of about $20\%$.

In order to determine the luminosity function for the 15 
contact binaries, we simply used 1-mag wide 
absolute-magnitude bins by dividing 
the (corrected) number of stars in each bin by the 
respective volume $V=\frac{4\pi}{3}(r_2^3-r_1^3)C$. 
$C$ denotes the fraction of the sphere covered by the 
survey and is equal to $1.09 \times 10^{-4}$ for the LMC 
and $5.82 \times 10^{-5}$ for the SMC. 
The results are given in table \ref{CB-LF-table}.

\begin{table}
\caption{The luminosity function for the 15 close contact 
binaries towards the LMC and the SMC. The limiting distances 
are determined from 
$d={\rm dex}{((V-M_V+5-A_V)/5)}$ 
with apparent magnitude limits 
$V \in (13 \pm 0.3,21 \pm 0.3)$. 
The errors in the number of stars $n$ are calculated 
according to the Poisson statistics, which contributes 
to the uncertainty in $\phi$ as the main factor. 
Furthermore, we arbitrarily set the number of stars 
equal to one for the SMC bin centred at $M_V=6$ with
zero star counts; an assumption which is consistent with the 
$1-\sigma$ Poisson error. }
\label{CB-LF-table}
\begin{tabular}{@{}ccccccc}
Cloud 	& $M_V$  & $r_1$ 	& $r_2$ 	& $n$ 	& $\phi$  \\
	&$(mag)$ & $(kpc)$	& $(kpc)$	& 	& $(\frac{stars}{kpc^3})$ 	\\
\hline
SMC	& 3	& 1.19	& 30.00	& 1.5	& $0.23 \pm 0.21$ \\
	& 4	& 0.75 	& 18.93	& 4.5	& $2.71 \pm 1.71$ \\
	& 5	& 0.48 	& 11.94	& 1.5	& $3.62 \pm 3.31$ \\
	& 6	& 0.30 	& 7.54	& 1.5 	& $14.39 \pm 13.18$ \\
\hline
LMC	& 3	& 1.13	& 28.41	& 1.6	& $0.15 \pm 0.14$ \\
	& 4	& 0.71 	& 17.93	& 4.8	& $1.82 \pm 1.12$ \\
	& 5	& 0.45 	& 11.31	& 8.0	& $12.10 \pm 6.59$ \\
	& 6	& 0.28 	& 7.14	& 1.6	& $9.63 \pm 8.60$ \\
\end{tabular}
\end{table}

\section[MS luminosity function]{Stellar density model 
and luminosity function for MS stars}

In order to establish the {\it relative\/} numbers of
contact binaries, in relation to MS stars, we attempted to
derive the predicted numbers of stars in the same search volumes
as those of the OGLE and MACHO surveys.
According to the pioneering work of \citet{gr83} and 
\citet{baso84}, the stellar distribution in the Milky 
Way can be modelled by a double exponential thin and 
thick disk and a spheroidal halo. 
Here, we describe the stellar density by such a three component
model
\begin{eqnarray*}
 n(z,R)=n_{thin}(z,R)+n_{thick}(z,R)+n_{sp}(z,R),
\end{eqnarray*}
where $(z,R)$ are the galacto-centric coordinates. The thick and the thin 
disks are modelled by
\begin{eqnarray*}
  n_{thin}(z,R)&=&n_0\cdot(1-c_{thick}-c_{sp})\\
		&\times&\exp{\left[\frac{-|z|}{z_{thin}}\right]}
		\exp{\left[-\frac{R-R_0}{h_{thin}}\right]} \\
  n_{thick}(z,R)&=&n_0\cdot c_{thick} \\
		&\times&\exp{\left[\frac{-|z|}{z_{thick}}\right]}
		\exp{\left[-\frac{R-R_0}{h_{thick}}\right]},
\end{eqnarray*}
where $z_{thin}$ and $z_{thick}$ denote the scale heights and
$h_{thin}$ and $h_{thick}$ the scale lengths for 
the thin and the thick disk respectively. Furthermore, $n_0$ is the 
local stellar density, and $c_{thick}$ and 
$c_{sp}$ are the local--density normalizations of the thick disk 
and the halo relative to the thin disk.
In this research, we use a projected de Vaucouleurs spheroid 
\citep{yo76,baso84} of the form
\begin{eqnarray*}
 n_{sp}(R^{\prime})&=&n_0\cdot c_{sp}\\
&\times&\exp\left[-10.093\left(\frac{R^{\prime}}{R_0}\right)^{\frac{1}{4}}+10.093\right]
\left(\frac{R^{\prime}}{R_0}\right)^{-\frac{7}{8}}.
\end{eqnarray*}
Here, $R^{\prime}$ is given by
\begin{eqnarray*}
 R^{\prime}=\left(R^2+\left(\frac{z}{\kappa}\right)^2\right)^{\frac{1}{2}},
\end{eqnarray*}
where $\kappa$ is the axis ratio of the de Vaucouleurs spheroid.

In our model, we used $z_{thin}=300 \pm 50 \ \mathrm{pc}$, 
$z_{thick}=1000 \pm 200 \ \mathrm{pc}$, 
$h_{thin}=2.5 \pm 1 \ \mathrm{kpc}$,  
$h_{thick}=3.5 \pm 1 \ \mathrm{kpc}$, 
$c_{thick}=0.05 \pm 0.03$, 
$c_{sp}=0.0015 \pm 0.001$, 
$\kappa=0.55 \pm 0.1$ and 
$R_0=8 \pm 0.5 \ \mathrm{kpc}$. 
These parameters were chosen in consideration of the work of
\citet{gr83}, \citet{kg89}, \citet{fm94}, 
\citet{bs97}, \citet{lm03} and \citet{ju05}. 
The rather large uncertainties in the
parameters reflect ranges in the results 
obtained in the mentioned papers. 

\begin{figure}
\begin{center}
\includegraphics[width=84mm]{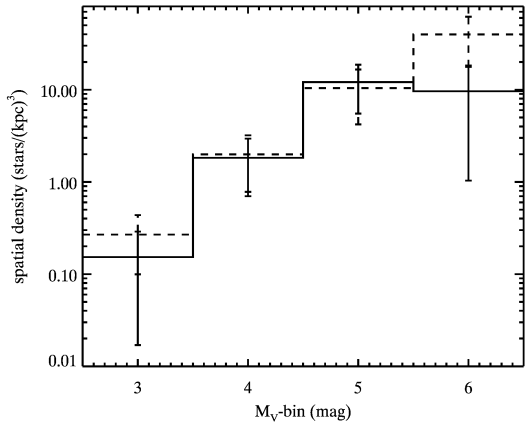}
\caption{\label{fig-lf_lmc}Luminosity function towards the LMC. 
The continuous line represents the contact binary luminosity function, 
whereas the dashed line represents the MS luminosity 
function divided by a factor of 650.}
\end{center}
\end{figure}

\begin{figure}
\begin{center}
\includegraphics[width=84mm]{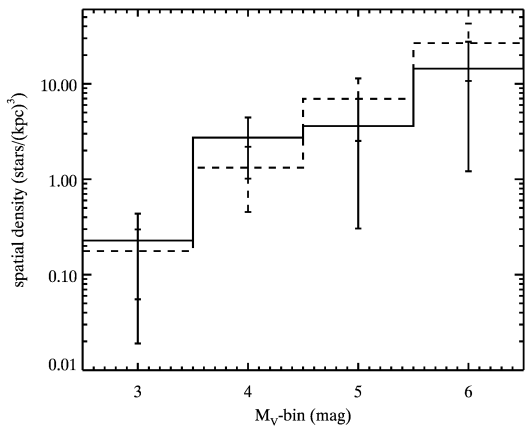}
\caption{\label{fig-lf_smc} Luminosity function towards the SMC. The
same as in figure \ref{fig-lf_lmc} except for the scaling factor 
of 600 instead of 650.}
\end{center}
\end{figure}

\citet{wi83} derived a widely accepted luminosity 
function for MS stars in the solar neighbourhood using a 
sample of nearby stars, which we used to determine the 
stellar density in the solar neighbourhood $n_0$ as a 
function of absolute magnitude. With those results and 
the stellar distribution $n(z,R)$ as described above, 
we calculated the mean stellar density of MS stars 
in each volume corresponding to the four, 1-mag wide,
absolute magnitude bins. By doing so, we were able 
to compare the derived contact binary luminosity 
function with the expected MS luminosity function 
for the same portion of the sky. The luminosity functions 
towards the LMC and the SMC are plotted in 
figures \ref{fig-lf_lmc} and \ref{fig-lf_smc} respectively. 
The contact binary luminosity functions were found 
to be best approximated by the MS luminosity function, 
if the latter was divided by $650$ and $600$ for the LMC and the SMC, 
respectively.

\section{Conclusions}

We conclude that each of the two samples yields 
the relative frequency of one contact binary among 
about $600$ MS stars in the two conical volumes towards 
the SMC and the LMC at galactic latitudes of $-44\degr$ 
and $-33\degr$ respectively. These estimates have uncertainties of 
roughly $50 \%$ due to the large uncertainties 
arising from small number statistics and to the still
uncertain parameters of the spheroidal component of the Galaxy.
This component has been explicitly accounted for 
in the previous section and is estimated to contribute 
to the total number of stars in the SMC
and the LMC search volumes at levels of about 30\% and 20\%, respectively. 
The contribution varies with the $M_V$. Specifically, for the SMC, it changes
from 38\% for the $M_V$-bin with the largest distances to 17\% for the
$M_V$-bin centred at 6 mag; for the LMC, it changes from 29\% to 10\% in
the same range of $M_V$.

The results on the relative frequency of contact binary
stars at large galacto-centric distances 
is consistent with estimates for the solar 
neighbourhood in previous work \citep{ru06}. These are the very 
first data available on the contact binary 
distribution at high galactic latitudes.

\section*{Acknowledgments}
This work was done during the stay of PDN
as an Exchange Undergraduate Student
at the Department of Astronomy and Astrophysics of
the University of Toronto. Thanks are expressed to
the members of the Department for hospitality and
support. PDN would also like to thank Bryce Croll and Mirza Ahmic 
whose suggestions on how to improve the paper have been most helpful. 
Thanks are due to the OGLE and MACHO teams for making their
results available for direct on-line access.
Support from the Natural Sciences and Engineering Council of Canada
to SMR is acknowledged with gratitude.

\end{document}